\documentclass[10pt]{article}
\usepackage{amssymb,amsmath,latexsym,amscd,amsfonts}
\usepackage{graphics}
\usepackage{epsfig} 
 
\newtheorem{thm}{Theorem}[section]

\newtheorem{deff}{Definition}[section] 
 
\newtheorem{claim}{Claim}[section] 
\newtheorem{lem}{Lemma}[section]

\newtheorem{corol}{Corollary}[section]

\newenvironment{proof}{{\noindent{\it Proof: }}}{$\Box$}

\def\calH{{\cal H}}

\def\C{\mathbb{C}} 
 
\def\mod{{\rm{mod}}}

\def\tr{{\rm{tr}}}

\def\ra{\rangle} 
\def\la{\langle}

\newcommand\ket[1]{{ |{#1} \rangle }} 
\newcommand\bra[1]{{ \langle {#1} | }} 

\newcommand{\dnote}[1]{{\bf (Dorit:} {#1}{\bf ) }} 
\newcommand{\ignore}[1]{} 
 
\newcommand{\onote}[1]{} 
\newcommand{\dfinalnote}[1]{}
\newcommand{\znote}[1]{}
 
\newcommand{\poly}{\rm{poly}}
 
\renewcommand{\epsilon}{\varepsilon}

\def\hpic #1 #2 {\mbox{$\begin{array}[c]{l}
\epsfig{file=#1,height=#2} \end{array}$}} 
\def\vpic #1 #2 {\mbox{$\begin{array}[c]{l}
\epsfig{file=#1,width=#2}\end{array}$}} 
 
\def\tl{TL_n(d)} 
\def\gtl{gTL_n(d)} 
\def\tr{\mbox{tr}}

\def\C{\mathbb{C}}

\def\TQFT{{\sf{TQFT}}}
\def\DNA{{\sf{DNA}}}
\def\hadamard{{\sf{Hadamard}}}
\def\BQP{{\sf{BQP}}}

 \def\sm2{\vspace*{-.02in}}
\begin{document}
%\conferenceinfo{STOC'06,} {May21--23, 2006, Seattle, Washington, USA.} 
%\CopyrightYear{2006} 
%\crdata{1-59593-134-1/06/0005}

% this is just before zeph sends me his remarks proof reading, 
% 18:00 before shortenning  18/3
\title{A Polynomial Quantum Algorithm for Approximating the Jones 
Polynomial} 

\ignore{
\numberofauthors{3}
\author{
\alignauthor Dorit Aharonov \\
\affaddr{Hebrew University} \\
\affaddr{Jerusalem} \\
\affaddr{Israel} \\
\email{doria@cs.huji.ac.il}
\alignauthor Vaughan Jones\\
\affaddr{U.C. Berkeley} \\
\affaddr{Berkeley} \\
\affaddr{California} \\
\email{vfr@math.berkeley.edu}
\alignauthor Zeph Landau  \\
\affaddr{The City College of New York} \\
\affaddr{New York} \\
\affaddr{New York} \\
\email{landau@sci.ccny.cuny.edu}}}

\author{ 
 Dorit Aharonov 
 \thanks{School of Computer Science and Engineering, 
 The Hebrew University, Jerusalem, Israel. 
 doria@cs.huji.ac.il.} 
%Research supported by ISF grant 032-9738 and Alon Fellowship} 
 \and Vaughan Jones \thanks{Department of Mathematics, U.C.Berkeley}
\and 
 Zeph Landau 
\thanks{Department of Mathematics, The City College of  New York, NY}}

\maketitle{}

\begin{abstract}
The Jones polynomial, discovered in $1984$ \cite{jones:poly},
is an important knot invariant in topology. 
Among its many connections to various mathematical and physical areas, 
it is known (due to Witten \cite{witten}) to be 
intimately connected to Topological Quantum Field Theory ($\TQFT$). 
The works of Freedman, Kitaev, Larsen and Wang 
\cite{tqftsimulation, tqftuniversal}
provide an efficient simulation of $\TQFT$ by a quantum computer,
and vice versa. These results implicitly 
imply the existence of an efficient quantum algorithm that 
provides a certain additive approximation of the Jones 
polynomial at the fifth root of unity, $e^{2\pi i/5}$, 
and moreover, that this problem is $\BQP$-complete. 
Unfortunately, this important algorithm was never explicitly 
formulated. Moreover, the results in 
\cite{tqftsimulation,tqftuniversal}
are heavily based on $\TQFT$, which makes the 
algorithm essentially inaccessible to computer scientists. 
 
We provide an explicit and simple polynomial quantum 
algorithm to approximate the Jones 
polynomial of an $n$ strands braid with $m$ crossings 
at any primitive root of unity $e^{2\pi i/k}$, where the running time 
of the algorithm is polynomial in $m,n$ and $k$. 
 Our algorithm is based, rather than on $\TQFT$,
on well known mathematical results (specifically, 
the path model representation of the braid group and the uniqueness
of the Markov trace for the Temperly Lieb algebra). 
By the results of \cite{tqftuniversal}, our algorithm solves a $\BQP$ 
complete problem. 

The algorithm we provide exhibits a structure which 
we hope is generalizable to other quantum algorithmic problems. 
Candidates of particular interest are the approximations of other 
downwards self-reducible $\#$P-hard problems, most notably, the 
important 
open problem of efficient approximation of 
the partition function of the Potts model, a model 
which is known to be tightly connected to the Jones 
polynomial \cite{jonespotts}.
\end{abstract}

\section{Introduction} 
Since Shor's breakthrough discovery in 1994 \cite{shor}, quantum 
algorithms 
with an exponential speedup over the best known classical 
algorithms have been shown for a number of problems 
(e.g., \cite{legendre,watrous,hallgren,kuperberg}). 
All these problems and algorithms share 
some common features: the problems 
are group or number theoretic in 
nature and the key component of each algorithm is 
the quantum Fourier transform\footnote{One interesting
 exception is the beautiful result of 
 \cite{childs} who used random walks 
to achieve an exponential algorithmic speed-up for an oracle graph 
problem. 
This technique, however, 
has not yet found applications.}. 
Arguably, the greatest 
challenge of quantum computation is the discovery of 
new algorithmic techniques. 

In this paper we describe a polynomial time quantum algorithm
that approximates the $\#$P-hard problem of evaluating 
the Jones polynomial at certain roots of unity. 
The best classical algorithm for this problem is exponential. 
Our algorithm is significantly different from all previously 
known quantum algorithms that achieve an exponential speed
 up in the following three important ways: 1) it solves a problem 
which is combinatorial rather than group or number theoretic 
in nature, 2) it does so not by using the Fourier transform, 
but instead, by exploiting a certain 
structure of the problem and encoding it into the nature of 
the unitary gates being used, and 3) it solves a problem that is 
BQP-hard \cite{freedman}, that is,  a problem that captures all the 
power 
of the quantum model.

The connection between quantum computation and
 the Jones polynomial was first made through the 
series of papers \cite{freedman0,freedman,
tqftsimulation,tqftuniversal}. A model of quantum computation 
based on Topological Quantum Field 
Theory ($\TQFT$) and Chern-Simons theory was 
defined in \cite{freedman0,freedman}, and Kitaev, Larsen, Freedman and 
Wang showed that this model 
is polynomially equivalent in computational power to the standard 
quantum computation model in \cite{tqftsimulation, tqftuniversal}. 
These 
results, combined with a deep connection 
between $\TQFT$ and the value of the Jones polynomial at particular 
roots 
of unity discovered by Witten $20$ years ago \cite{witten} , implicitly 
implies an 
efficient quantum algorithm for the approximation of the 
Jones polynomial at the value $e^{2\pi i/5}$. 
This connection is also discussed, from the point of view of $\TQFT$, 
in Preskill's notes \cite{preskill}. 
Unfortunately, 
the important quantum algorithm implied by these intriguing 
results, though referred to in
\cite{lovasz}, was never explicitly formulated. 

In this paper we use a different route to connect quantum 
computation and the Jones polynomial, 
 one that does not involve  $\TQFT$.
We present an explicit and simple to state algorithm 
for the above problem, which is 
based purely on algebraic results from more than $20$ years ago. 
\ignore{This provides 
an algorithm for this important problem which 
is explicit and understandable in combinatorial language.} 
Moreover, our algorithm works for all roots of unity of the 
form $e^{2\pi i/k}$, going beyond the discussions 
in previous works involving only constant $k$'s. 
We now describe the precise problem that we solve. 

\subsection{Background on the Jones Polynomial} 
A central issue in low dimensional topology is that of 
{\it knot invariants}. A knot invariant is a function on knots (or
links --i.e. circles embedded in $R^3$) which 
is invariant under isotopy of the link, i.e., it 
does not change
under stretching, moving, etc., but 
no cutting. In $1984$, Jones \cite{jones:poly}
discovered a new knot invariant, now called 
 the Jones polynomial $V_L(t)$, which is a Laurent polynomial in 
$\sqrt t$ with integer coefficients, and which
is an invariant of the link $L$.
In addition to the important role it has played in low dimensional 
topology, the Jones polynomial has found
applications in numerous fields, from $\DNA$ recombination \cite{cozz}, 
to statistical physics \cite{jonespotts}. 

The Jones polynomial can also be defined as a function of braids. 
A braid of $n$ strands and $m$ crossings 
is described pictorially by $n$ strands hanging
alongside each other, with $m$ crossings, each of two adjacent
strands. 
A braid may be ``closed'' to form a link by tying its ends together. 
In this paper we will be interested in two ways to perform such 
closures, 
namely, the {\it trace closure} and the {\it plat closure} 
%of the braid 
(to be defined 
later). We will be interested in the Jones polynomial of links that are 
trace or plat closures of braids. 

The essential aspect of the Jones polynomial of a link $L$ can be 
computed by the following process: project $L$ to the
 plane keeping track crossings to get what is called a {\it link 
diagram}. 
Now replace 
every crossing $\hpic{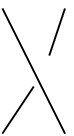} {.3in} $ by a particular 
linear combination (with coefficients being
 functions of the parameter $t$) of the 
pictures $\hpic{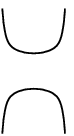} {.3in} $ 
 and $\hpic{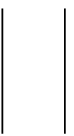} {.3in} $. 
The result is 
%a linear combination of links with no 
%crossings, i.e., 
a linear combination of diagrams containing only 
closed loops. Replace 
each of these diagrams with a particular 
function of the parameter $t$ and the number of loops. The resulting 
function of $t$ is a scaled version of the Jones polynomial $V_L(t)$. 

From the moment of the discovery of the Jones polynomial, 
the question of how hard it is to compute was important. 
There is a very simple inductive algorithm (essentially due to Conway  
\cite{Conway}) to compute it by changing
crossings in a link diagram, but, naively applied, this takes 
exponential time in the number of crossings. It was shown \cite{Welsh} 
that the 
computation of $V_L(t)$ is $\#$P-hard for all but a few values 
of $t$ where $V_L(t)$ has an elementary 
interpretation.
Thus a polynomial time algorithm for computing $V_L(t)$ for any 
value of $t$ other than those elementary ones is unlikely. 
Of course, 
the $\#$P-hardness of the problem does not rule out the possibility of 
good approximations; see, e.g.,  \cite{permanent}. 
Still, the best classical algorithms to approximate the 
Jones polynomial at all but trivial values are exponential. 
\subsection{Our Results} 
We show an efficient, explicit, and simple 
quantum algorithm to approximate the 
Jones polynomial at all points of the form $t=e^{2\pi i/k}$.
It will in fact be easier to use the parameter $A=t^{-1/4}$. 
We prove the following for the trace closure $B^{tr}$ and the plat 
closure $B^{pl}$ of a braid $B$:

\begin{thm}\label{thm:maintrace} 
For a given braid $B$ with $n$ strands and $m$ crossings, and a given
integer 
$k$, there is a quantum algorithm which is polynomial in $n,m,k$
which with all but exponentially small probability,
outputs a complex number $r$
 with $|r-V_{B^{tr}}(e^{2\pi i/k})|< \epsilon d^{n-1}$
%$\in [V_{B^{tr}}(e^{2\pi i/k})-\epsilon d^{n-1},V_{B^{tr}}
%(e^{2\pi i/k})+\epsilon d^{n-1}]$ 
where $d=-A^2-A^{-2}$, and $\epsilon$ is inverse polynomial in $n,k,m$. 
\end{thm} 

\begin{thm}\label{thm:mainplat} 
For a given braid $B$ with $n$ strands and $m$ crossings, and a given
integer 
$k$, there is a quantum algorithm which is polynomial in $n,m,k$
which with all but exponentially small probability,
outputs a complex number $r$ with $|r-V_{B^{pl}}(e^{2\pi i/k})|< 
\epsilon d^{3n/2}/N$
%$\in [V_{B^{pl}}(e^{2\pi i/k})-\epsilon 
%d^{3n/2}/N,{\cal R}eV_{B^{pl}}
%(e^{2\pi i/k})+\epsilon d^{3n/2}/N]$ 
where $d=-A^2-A^{-2}$ and $\epsilon$ is inverse polynomial in $n,k,m$
($N$ is an exponentially large factor to be defined 
later). 
\end{thm}

We remark that the approximation we provide here is 
additive, namely the result lies in a given window, whose 
size is independent of the actual value we are trying to 
approximate. This of course is much weaker than a 
multiplicative approximation, which 
is what one might desire (see discussion in \cite{lovasz}). 
One might wonder if under such weak 
requirements, 
the problem remains meaningful at all. It turns out that, in fact, 
this additive approximation problem is hard for quantum 
computation:
\begin{thm} {\bf Adapted from 
Freedman, Larsen and Wang \cite{tqftuniversal}}
  \label{thm:universal} 
The problem of approximating the Jones polynomial of the plat closure 
of a braid at $e^{2\pi i/k}$ for constant $k$, 
to within the accuracy given in Theorem \ref{thm:mainplat}, 
is $\BQP$-hard. 
\end{thm}
This result was recently strengthened by Aharonov and 
Arad \cite{arad} to any $k$ which is
 polynomial in the size of the input, 
namely, for all the plat closure 
cases for which our algorithm is polynomial in the size of the braid. 

Curiously, the hardness results of \cite{tqftuniversal,arad} 
are not known to hold (regardless of $k$)
 for the approximation of the 
trace closure for which we give an algorithm as well. 
We discuss the difference between the two problems 
in the open questions section.

\ignore{We remark that the approximation we provide here is additive, 
 and that this cannot be done to within arbitrary precision; 
the exact scale is important. \dnote{A multiplicative 
approximation, which is much harder to achieve, 
is ruled out since this would ... fill in}.}

\subsection{Description of the Algorithm}
The essence of the algorithm lies in 
the fact that for braids with $n$ strands, 
the pictures that one gets by replacing 
each crossing by one of the two pictures 
 $\{ \hpic {capcup.eps} {.3in} , \ \hpic{id.eps} {.3in} \}$, can be 
assigned the structure of an algebra. This algebra is
 called the Temperley Lieb algebra, 
and denoted by $TL_n$. In fact, 
the map from the crossing to the appropriate 
linear combination of the above two pictures defines a 
{\it representation} of the group $B_n$ of braids of $n$ strands, 
 inside the $TL_n$ algebra. 
The Jones polynomial of the trace 
closure of a braid can be seen as a certain trace function 
 (i.e., a linear function satisfying $tr(AB)=tr(BA)$) on the image 
of the braid in the $TL_n$ algebra. 

Our goal is then to design an algorithm that approximates this trace. 
 To this end we use an important fact about this trace: it
satisfies an additional property 
called the {\it Markov property}. Moreover, 
this property makes it unique; any trace 
function on the $TL_n$ algebra (or a representation of it) that 
satisfies this property is equal to the above trace!
%To this end we use a very useful fact about this trace: 
%it satisfies an additional property called the Markov property, 
%that makes  it unique, namely, 
%any other trace function on the $TL_n$ algebra 
%(or a representation of it) that satisfies this property is 
%equal to the above trace! 
This leads us to the key idea of the 
algorithm: 
suppose we can define a representation of the $TL_n$ algebra 
by matrices operating on qubits, and we can identify and estimate 
 the trace that satisfies the Markov property on 
this representation. Then by the uniqueness of this trace 
we can estimate the Jones polynomial.

But what is the representation that should be used? 
If our intent is to design a quantum algorithm, it is best if the 
representation induced on the braid group be unitary, so that 
we can hope to approximate its trace by a quantum computer. 
Fortunately, it is in fact possible to give representations
of the Temperley Lieb algebra which induce {\it unitary} 
representations of the braid group. 
Such representations were constructed in
\cite{jones:hecke,jones:poly} and 
are called the {\it path model} representations. 
If we want to evaluate the 
Jones polynomial $V_L(t)$ for $L$ a closure of a braid in $B_n$, 
and $t=e^{2\pi i/k}$, we use the 
$k$th path model representation of $B_n$. 
It is fairly straight forward to adapt these representations 
to work on the space of $n$ qubits, and moreover, to show that 
the resulting unitary matrices (namely, the images of the generators 
of the braid group) can be applied efficiently by a quantum computer. 
We find that the image (by the path model representation) of an
entire braid $B$ can be applied efficiently by 
 a quantum computer.
Let us call the unitary matrix
corresponding to a braid $Q(B)$.

To approximate the Jones polynomial of a trace closure of the braid, 
it suffices to approximate the Markov trace of $Q(B)$. 
This is done using standard quantum and classical algorithmic 
techniques, 
including the well known $\hadamard$ test. 
The algorithm for the plat closure builds on similar ideas, 
though it is not directly stated in terms of traces. 
Thus we obtain a polynomial quantum algorithm for the $\BQP$-complete 
problem of approximating the Jones polynomial of a plat closure of a 
braid. 

We remark that after we completed this 
work, we learned about a previous 
independent attempt to prove similar results
\cite{subramaniam}. 
% using unitary representations of the braid group. 
Unfortunately, the work of \cite{subramaniam} is 
greatly flawed, and in particular claims to 
provide an exact solution to the $\#P$-hard problem.

\subsection{Conclusions and Further Directions} 
We have provided a simple algorithm for a $\BQP$-complete problem, 
which is different in its methods than previous quantum algorithms. 
In essence, what it does is to isolate a certain local structure of the 
problem, and assign gates which somehow exhibit the same local 
structure. 
Our hope is that this more combinatorial direction in quantum 
algorithms will lead to further progress in the area.

In particular, one very interesting related question 
is an important problem from mathematical physics:
 that of approximating the partition function
of the Potts model \cite{pottshard}, which  
 is known to be tightly connected to the Jones polynomial 
\cite{jonespotts}, 
and its exact evaluation is once again
$\#$P-hard \cite{pottshard}. 
We hope that the results of this paper will lead to
progress in this question, or in other questions related to
approximating $\#P$-complete problems. 

\ignore{
Another open question is that of the computational complexity of the 
approximation of the Jones polynomial of the plat closure of a braid, 
at points $e^{2\pi i/k}$ for $k$ which grows polyhnomially 
in the number of strands. We believe that these problems are also 
$\BQP$-complete, but we were unable to prove this so far. The issue 
here is 
that in order to approximate any quantum circuit by the braid 
generators, one needs to apply the well known Solovay-Kitaev 
theorem \cite{nielsen}, but this does not apply to 
non-constant generators as we get for non-constant $k$.} 

We briefly discuss the relation between the 
plat and the trace closures problems. 
It is known that any plat closure of 
a braid can be transformed efficiently into a trace closure of some 
other 
braid \cite{vogel}. The reader might therefore find it curious that one 
of these problems is $\BQP$-complete, while the other one is 
not known to be
so. The explanation lies in the fact that the quality of the 
approximation in both algorithms depends {\it exponentially} on
the number of strands in the braid. The transformation from plat to 
trace 
closures requires, in the worst case, a significant increase in
 the number of strands. This, unfortunately, degrades 
the quality of the approximation exponentially. 
The computational complexity of the trace closure problem is left open.

Finally, we believe that this 
paper helps to clarify and demystify (at least one direction of) the 
intriguing equivalence between quantum computation and 
the problem of approximating 
the Jones polynomial. We hope
  this connection leads to a deeper understanding of quantum 
computation complexity. 
\ignore{zeph, should we add this? 
That this connection may lead to a deeper understanding 
of quantum computation 
 is demonstrated by the following simple but interesting 
example; It is an implication of the direction of 
the Jones Polynomial - Quantum computation 
equivalence not given in the work, namely 
of Theorem \ref{thm:universal}. 
It is well known that the computation of the Jones polynomial 
of links can be calculated in time at most exponential in the tree 
width of 
the link \cite{}. Theorem \ref{thm:universal} can thus be used to 
give an efficient classical simulation of quantum circuits 
of small tree width. This is a simple way to derive the recent result 
of 
\cite{shi}.} 

{~}

\noindent{\bf Organization of paper:}
Section \ref{sec:background} provides the background and the necessary 
definitions, starting from quantum computation, the $\hadamard$ test, 
the braid group, algebras and representations, the Jones polynomial, 
the Temperley Lieb algebras and the path model representation. 
Using these notions we describe the algorithms in
 Section \ref{sec:alg} and prove their correctness. 

\section{Background}\label{sec:background} 

\subsection{Quantum Computation} 

For background on quantum computation, see \cite{nielsen}.
We merely mention here that, strictly speaking, we use 
here a quantum-classical hybrid model of computation, in which a 
classical 
probabilistic Turing machine performs calls to a quantum computer, 
and uses its outcomes to perform some classical computations. 
It is standard that this model can be simulated efficiently 
by the standard quantum computation model.

\subsection{The $\hadamard$ Test}
The following fact 
is standard in quantum computation. 
If a state $|\alpha\ra$ can be generated efficiently, 
and a unitary $Q$ can be applied efficiently, 
then there exists an efficient quantum circuit whose output is 
 a random variable $\in \{-1,1\}$, and whose expectation is 
${\cal R}e\la \alpha|Q|\alpha\ra$. 
%This is very similar to the usual ${\sf SWAP}$ test 
%often used in quantum computation. 
Start with the two-register state
 $\frac{1}{\sqrt{2}}(|0\ra+|1\ra)\otimes|\alpha\ra$, apply  
$Q$ conditioned on the first qubit to get the state
$\frac{1}{\sqrt{2}}(|0\ra\otimes 
|\alpha\ra+|1\ra\otimes Q|\alpha\ra$,  
apply a $\hadamard$ gate on the first qubit, and measure. Output 
$1$ if the measurement result is $|0\ra$, $-1$ if the measurement 
result is $|1\ra$.  
The expectation of the output is exactly ${\cal R}e\la 
\alpha|Q|\alpha\ra$.
To get a random variable whose expectation is the imaginary part, 
start with the state 
$\frac{1}{\sqrt{2}}(|0\ra-i|1\ra)\otimes|\alpha\ra$ instead. 
\ignore{
and apply instead of the $\hadamard$ gate, the gate which takes $|0\ra$
to $\frac{1}{\sqrt{2}}(|0\ra+i|1\ra)$, 
and $|1\ra$ to $\frac{1}{\sqrt{2}}(|0\ra-i|1\ra)$.}

\subsection{Algebra Background}\label{app:algebra} 
An algebra is a vector space with a multiplication. The
multiplication
must be associative and distributive.
A representation
of a group $G$ inside an algebra
is a group homomorphism $\rho$ 
from $G$ to the group of invertible elements in the algebra, namely,
we require $\rho(g_1)\rho(g_2)=\rho(g_1g_2)$ for any 
$g_1,g_2\in G$. 

\ignore{once again, we do not need this terminology here.
The class function thus obtained using
the trace on matrices is called the {\it character} of the
representation. }
 
We say a representation is reducible if there exists a proper
subspace of vectors which is invariant under the group action.
If there is no such subspace, we say the representation is
irreducible.
 
We shall sometimes refer to a representation of G without specifying 
the 
algebra; in these cases we shall mean a representation inside the 
algebra
of $n\times n$ matrices. 
%We will also talk about the representation of an algebra. 
%The definition is similar to that of a representation of a group: 
We shall be interested in algebra representations as well: 
\begin{deff}
An $r$ dimensional representation $\Phi$ of an algebra
is a linear mapping from the algebra into the 
set of $r\times r$ complex matrices $M_r$, such that 
for any two elements $X,Y$ in the algebra, 
$\Phi(X)\Phi(Y)=\Phi(XY)$ .
\end{deff}
 
If a group is represented inside an algebra then any representation
of the algebra gives a representation of the group by composition.
 
Often, an algebra or a group is defined using a set of generators
and relations between them. In this case, a representation may be
defined 
by specifying the images of the generators, provided 
the same relations hold for the images as for the generators.

\subsection{The Braid Group}\label{sec:braid}
Consider two horizontal bars, one on top of the other, with n pegs on
each.
 By an {\it n strand braid} we shall mean a set of n strands such
that: (1) Each strand is tied to one peg on the top bar and one peg on
the bottom bar, (2) Every peg has exactly one end attached to it, 
(3) The strands may pass over and under each other, 
(4) The tangent vector of every strand at any point along the path
from top to bottom always has a non-zero component in the
downward direction. 
Here is an example of a 4-strand braid:
\[ \vpic {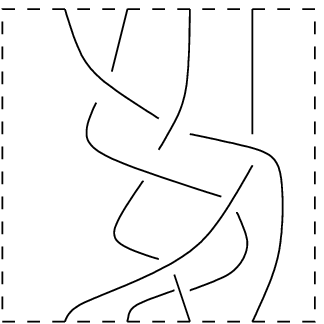} {.7in} . \] 
The set of n-strand braids, $B_n$, has a group structure with
multiplication as follows. Given two n-strand braids $b_1, b_2$,
place braid $b_1$ above $b_2$, remove the bottom $b_1$ bar and the
top $b_2$ bar and fuse the bottom of the $b_1$ strands to the top of
the $b_2$ strands. 
 
The product of the above 4-strand braid with the 4-strand braid
$\vpic {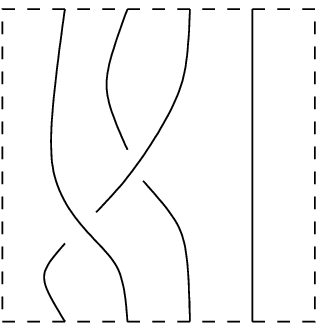} {.5in} $ is: 
\[ \vpic {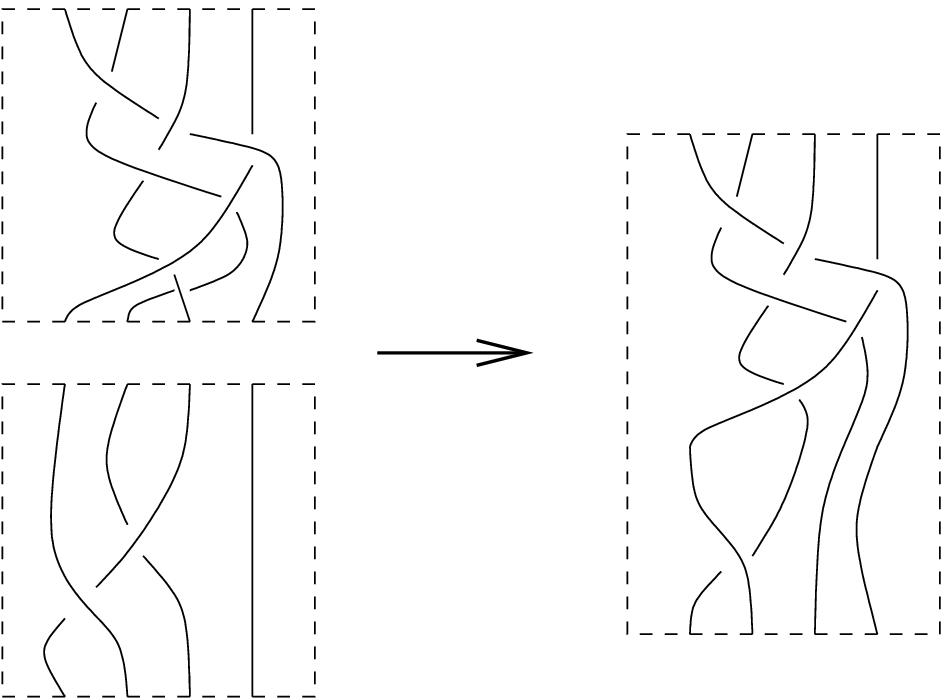} {2in} . \]

An algebraic presentation of the braid group due to Artin is as
follows \cite{Art}: Let $B_n$ be the group
with generators $\{ 1, \sigma_1, \dots \sigma_{n-1} \}$ and
relations 
\begin{enumerate} 
\item $\sigma_i\sigma_j= \sigma_j \sigma _i $ for $ |i-j|\geq 2$, 
\item $\sigma_i \sigma_{i + 1}\sigma _i= \sigma_{i + 1}\sigma_i
\sigma _{i+ 1}$. 
\end{enumerate}
 
This algebraic description corresponds to the pictorial picture of 
braids: 
$\sigma_i $ corresponds to the pictorial braid
$ \vpic {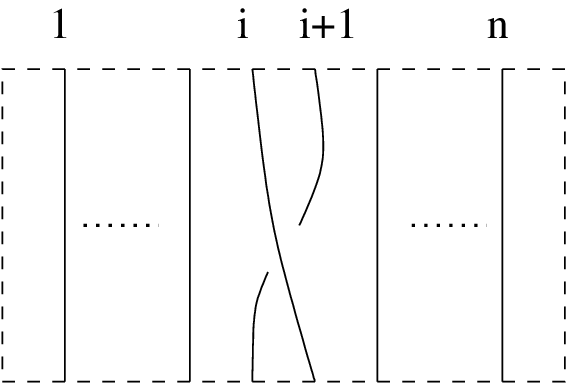} {.8in} $, and concatenating such pictures gives
a general braid in $B_n$.

\subsection{The Temperley-Lieb Algebras}\label{sec:tl} 
\begin{deff} \label{d:1} 
Given $n$ an integer and $d$ a complex number we define the
Temperley-Lieb algebra $\tl$ to be the algebra generated by $\{1,
E_1, \dots , E_{n-1}\}$ with relations
\begin{enumerate} 
\item $E_iE_j= E_j E_i$, $|i-j|\geq 2$, 
\item $E_i E_{i\pm 1} E_i= E_i$,
\item $E_i ^2=d E_i$. 
\end{enumerate} 
 \end{deff} 
\ignore{
When $d$ is real, as it will be throughout the paper, we define 
the involution on $\tl$, denoted by $*$, 
by 
$(E_{i_1}E_{i_2}\cdots E_{i_r})^*= E_{i_r}E_{i_{r-1}}\cdots E_1$, 
and extend by linearity.} 

There is a well known geometric description of $\tl$
due to Kauffman \cite{kauffman}.
It uses the notion of Kauffman
$n$-diagrams, which is best explained by an example, e.g.,
a Kauffman $4$-diagram: 
 \[ \vpic {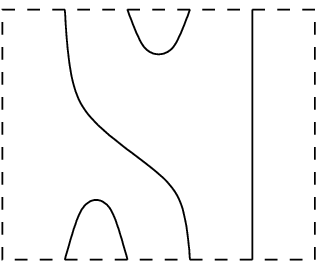} {1in} \] 
In general, a Kauffman $n$-diagram
is a diagram as above, with $n$ top pegs and $n$ bottom pegs, 
and no crossings and no loops. 
More formally: 

\begin{deff} 
Let $D_n$ be a rectangle with $n$ marked points on the top of the
boundary and $n$ marked points on the bottom. A {\it
Kauffman n-diagram} is a picture sitting inside $D_n$ consisting
of $n$ non-intersecting curves that begin and end at distinct marked
boundary points. We will consider two such diagrams equal if they
are isotopically equivalent (keeping the boundary fixed). 
\end{deff}

We define a vector space over these diagrams: 

\begin{deff} 
Let ${\cal K}_n$ be the vector space of complex linear combinations of
Kauffman n-diagrams. 
\end{deff} 
Multiplication of two Kauffman n-diagrams is done just like in the case 
of 
braids. 
To multiply a diagram $k_1$ with a diagram $k_2$ we stack 
$k_1$ on top of $k_2$, and fuse the matching ends of the strands.  The 
resulting diagram is a new Kauffman diagram $k_3$ with possibly some 
extra closed loops.  We define the product of $k_1$ and $k_2$ to be $d^m 
k_3 \in {\cal K}_n$ where $m$ is the number of extra closed loops.
%However, the resulting diagram may contain loops, which means it is not 
%in ${\cal K}_n$. Hence, the loops are removed, but the resulting 
%diagram 
%is multiplied by $d^m$ (where $m$ is the number of loops removed). 
For example, 
a multiplication of the above Kauffman 4-diagram with the Kauffman
4-diagram 
$\vpic {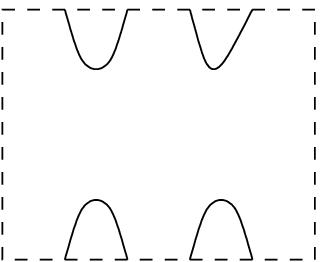} {.7in} $ results in: 
 
\[ \hpic {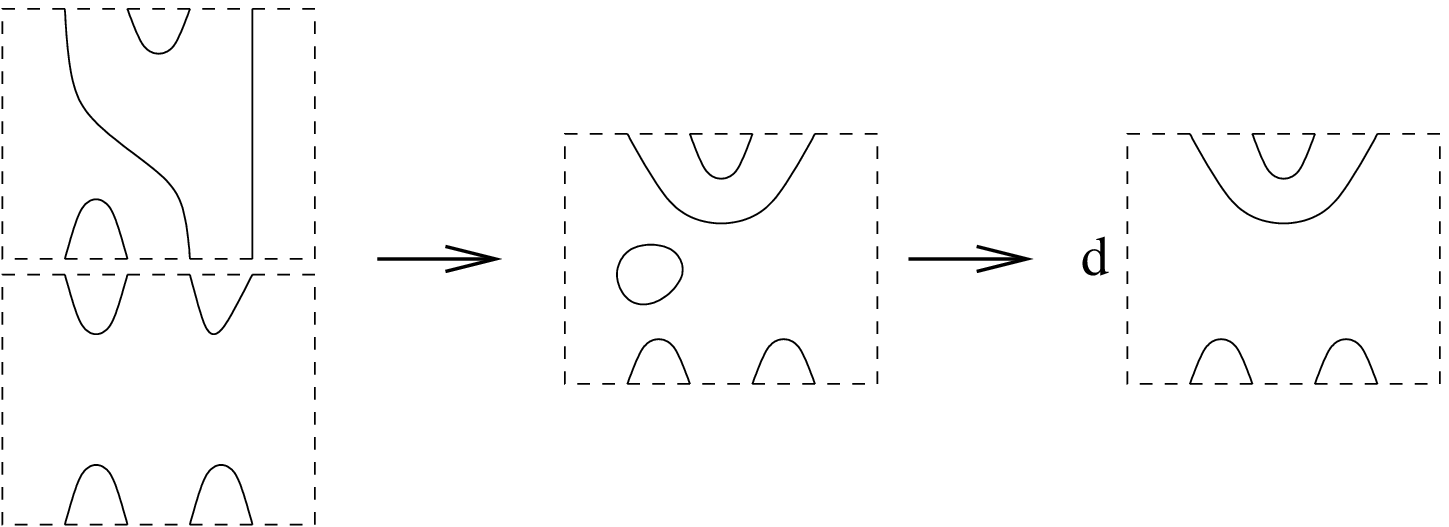} {1.4in} \] 

%This gives a multiplication on the vector space ${\cal K}_n$, 
%and thus, we have an algebra, denoted $\gtl$. 
This multiplication rule can be extended linearly to ${\cal K}_n$, we 
call the resulting algebra $\gtl$.

\ignore{More formally: 

\begin{deff} Let $\gtl $ denote the algebra consisting of the vector
space ${\cal K}_n$ with multiplication of two Kauffman diagrams $k_1$ 
and
$k_2$ as follows: 
 
Stack $k_1$ on top of $k_2$ so that the bottom $n$ marked points
points of $k_1$ align with the top $n$ marked points of $k_2$.
Connect the strands of $k_1$ ending at the bottom of $k_1$ to the
corresponding strands of $k_2$ starting at the top of $k_2$. The
resulting picture may have some closed loops; let $m$ be the number
of such loops. Erasing the closed loops yields a Kauffman diagram
$k_3$ with boundary the top of $k_1$ and the bottom of $k_2$. Define
the product $k_1 \cdot k_2= d^{m}k_3$. 
 
For a Kauffman n-diagram $k$, define $k^*$ to be the Kauffman
n-diagram that is the reflection of $k$ along a horizontal line.
Extend the $^*$ map conjugate linearly to all of $\gtl$. 
\end{deff}} 

The algebras $\tl$ and $\gtl$ are isomorphic: 
\begin{thm} 
The map $\psi: \tl \rightarrow \gtl$ given by the homomorphic
extension of $\psi (E_i)= \vpic {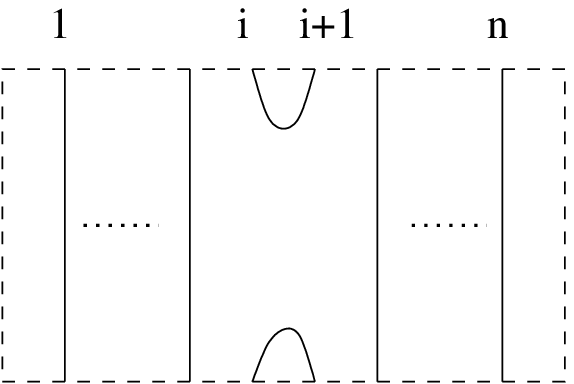} {1in} $ is an
isomorphism. 
\end{thm} 

\begin{proof}
 It is a simple and fun exercise to check that the image of the 
relations
given in Definition \ref{d:1} are relations in $\gtl$. For the 
remaining 
details see \cite{bischjones}. 
\end{proof}

We shall refer to the pictures of the form 
$\hpic{capcup.eps} {.3in} $, which generate $gTL_n(d)$, as
{\it capcups}. 

\medskip

\ignore{
It is clear that the subalgebra of $\gtl$ consisting of linear
combinations of only those Kauffman $n$ diagrams with a vertical
rightmost strand (i.e. diagrams where the top right marked point is
connected to the bottom right marked point by a vertical time) is 
isomorphic to
$gTl_{n-1}(d)$. We see then that there is a natural nesting of the
algebra inclusions $gTl_1(d) \subset gTL_2(d) \subset \dots \subset
gTl_n(d)$. These are exactly the image by the map $\psi$ 
of the natural algebra
inclusions $Tl_1(d) \subset TL_2(d) \subset \dots \subset Tl_n(d)$. }

\subsection{Representing $B_n$ Inside $TL_n(d)$} 

We define a mapping from the braid group to $\tl$:
%the Temperley-Lieb algebra: 
\begin{deff} \label{def:embed}
$\rho_A: B_n \mapsto TL_n(d)$ is defined by 
$$\rho_A(\sigma_i)= AE_i+A^{-1}I.$$ 
\end{deff} 

\begin{claim}\label{cl:embed} 
For a complex number $A$ which satisfies $d=-A^2-A^{-2}$, 
the mapping $\rho_A$ is a representation of the
 braid group $B_n$ inside $TL_n(d)$. 
\end{claim}

\begin{proof}
We need to check that the relations of the braid group are
satisfied by this mapping. 
 For $|i-j|>1$, 
$\rho_A(\sigma_i)$ commutes with $\rho_A(\sigma_j)$ since 
$E_i$ commutes with $E_j$.  
To show that
$\rho_A(\sigma_i)\rho_A(\sigma_{i+1})\rho_A(\sigma_{i})=
\rho_A(\sigma_{i+1})\rho_A(\sigma_{i})\rho_A(\sigma_{i+1})$, 
substitute to get an expression in $E_i$'s. 
Opening up the first expression we get
$A^3E_iE_{i+1}E_i+
AE_{i+1}E_{i}+AE_i^2+A^{-1}E_i+AE_iE_{i+1}+A^{-1}E_{i+1}+A^{-1}E_i+A^{-3}$.
The second expression gives 
$A^3E_{i+1}E_{i}E_{i+1}+
AE_{i}E_{i+1}+AE_{i+1}^2+A^{-1}E_{i+1}+AE_{i+1}E_{i}+A^{-1}E_{i}+A^{-1}E_{i+
1}+A^{-3}$. 
We remove similar terms, and 
using the relations of the $TL_n(d)$ it remains to 
show that $(A^{-1}+Ad+A^{3})E_{i}=(A^{-1}+Ad+A^{3})E_{i+1}$. 
This holds because the constants are $0$ due to the 
relation between $d$ and $A$. 
 \end{proof}

\ignore{From now on we fix $d=-A^2-A^{-2}$.}

\subsection{Unitary Representation of 
$B_n$}\label{sec:unitary}
Given a representation $\tau$ of $TL_n(d)$, 
we may use the representation of 
the braid group inside the $TL_n(d)$ 
algebra (Definition \ref{def:embed}) 
to derive a representation of $B_n$ by composition, as follows. 
Define the map $\varphi$ by specifying its operation 
on the generators $\sigma_i$ of $B_n$ 
to be $\varphi(\sigma_i)=\varphi_i=\tau(\rho_A(\sigma_i))= 
A\tau(E_i)+A^{-1}I$. 
This representation is unitary under certain constraints: 

\begin{claim}\label{cl:braidrepunitary} 
If $|A|=1$ and $\tau(E_i)$ are Hermitian for all $i$, 
then the map $\varphi$ is a unitary representation of $B_n$. 
\end{claim} 

\begin{proof}
$\tau(\rho_A(\sigma_i))\tau(\rho_A(\sigma_i))^\dagger
=(A^{-1}I+A\tau(E_i))((A^{-1})^*I+A^*\tau(E_i)^\dagger)
=I+A^{-2}\tau (E_i)+A^2\tau(E_i)+d\tau(E_i)=I$. 
\end{proof}

\subsection{Tangles} 
For this paper, we define a tangle to be a braid in which some 
of its crossings have been replaced by a 
picture of the form 
 $\{ \hpic {capcup.eps} {.3in} \}$. Braids and Kauffman diagrams 
are tangles.

\subsection{From Braids to Links} 

We can connect up the endpoints of a braid in a variety of ways to 
get links. We single out two such ways: 
\begin{deff} The {\it trace closure} of a braid $B$ 
 shall be the link achieved by connecting the
 strand at the rightmost top peg, around to the right,
 to the strand at the rightmost bottom peg, then connecting in the same 
way 
the next to rightmost top and bottom strands and so on.  
We denote the resulting link by $B^{tr}$. 
\end{deff}
\begin{deff} The {\it plat closure } of 
a $2n$-strand braid shall be the link formed by
connecting pairs of adjacent strands (beginning at the leftmost 
strand), 
on both the top and bottom. We denote the resulting link by $B^{pl}$. 
%connecting on the top (respectively on the bottom) the strands
%beginning at the odd numbered pegs (respectively the strands ending
%at the odd numbered pegs) to the neighboring peg immediately to the
%right.
\end{deff} 
 
Examples of the trace closure and the plat closure of the same
 4-strand braid are:
\[ \vpic {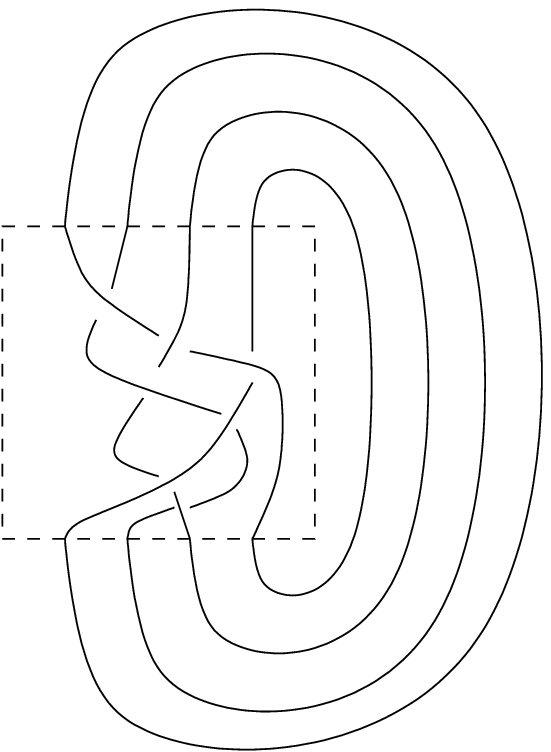} {1in}, {~} {~} {~}
 \vpic {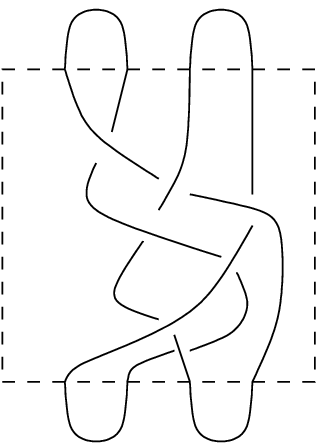} {1in} .\] 

These closures are also well defined for tangles. 

\subsection{The Jones Polynomial} \label{sec:kauffman}
A definition of the Jones 
polynomial $V_L(t)$ due to Kauffman \cite{kauffman} is as follows. 
We start by defining the Kauffman bracket $<L>$ as a polynomial in $A$ 
for $A$ such that $A^{-4}=t$. 

\begin{deff}
Consider a link $L$, given by a link diagram. 
A state $\sigma $ of $L$ shall
mean a choice, at each crossing $\hpic {crossing.eps} {.3in} $ 
of $L$, from the 
set $\{ \hpic {capcup.eps} {.3in}, \ \hpic{id.eps} {.3in} \}$. 
To
a state $\sigma $ of a link $L$ we associate the following
expression
 $\sigma (L)$:
 Let $\sigma ^+$ (respectively $\sigma ^-$) be the number of
crossings for which $\sigma$ chooses $\hpic {capcup.eps} {.3in} $
(respectively $ \hpic {id.eps} {.3in} $ ).
 Let $|\sigma|$ be the number of closed loops in the diagram
gotten by replacing each crossing $\hpic {crossing.eps} {.3in} $ by
the choice indicated by the state $\sigma$. 
 Define
\( \sigma (L) = A^{\sigma^+- \sigma ^-} d^{|\sigma|-1}. \)
The Kauffman bracket polynomial, also called the bracket state sum, 
for $L$, is defined to be 

\[ <L> = \sum_{\mbox{all states } \sigma} \sigma(L). \]
\end{deff}

To define the Jones polynomial, we consider 
oriented links, namely links with one arrow on each connected 
component. 
The connection between the Jones polynomial and the 
Kauffman bracket is given by a notion called the {\it writhe}:  
%defined on oriented links: 
\begin{deff}
For an oriented link $L$, assign to 
each crossing that looks like this $ \hpic {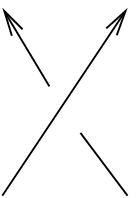} {.3in} $ 
the value $+1$, and to each crossing that
looks like this: $ \hpic {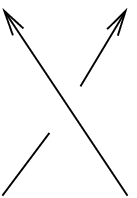}
 {.3in} $ the value $-1$.
The writhe of $L$ is the sum over all the crossings
of these signs.
\end{deff} 
 
\begin{deff}\label{def:jones}
The Jones polynomial of an oriented link $L$ is defined to be 
\[ V_L(t)= V_L(A^{-4})=(-A)^{3w(L)}\cdot <L>\]
where $w(L)$ is the writhe of the oriented link $L$,
and $<L>$ is the bracket state sum of the link $L$, ignoring 
the orientation. 
\end{deff}
Thus, the Jones polynomial is a scaled version
of the bracket polynomial. Moreover, 
the writhe of a link can be easily calculated from the 
link diagram, and hence 
the problem of calculating the bracket sum polynomial is equivalent 
in complexity to that of calculating the Jones polynomial. 

\subsection{The Markov Trace}

\begin{deff}
A linear function from an algebra to the complex numbers 
is called a {\it trace} if it satisfies 
$\tr (XY)=\tr(XY)$
for every two elements $X,Y$ in the algebra. 
\end{deff}
 
We define the following trace on $\gtl$. 

\begin{deff} 
The {\it Markov trace} $\tr: \gtl \rightarrow \C$ is defined on a
Kauffman n-diagram $K$ as follows.
 Connect the top $n$ labeled points to the bottom $n$ labeled
points of $K$ with non-intersecting
 curves, as in the trace closure.  Let $a$ be the number
 of loops of the resulting diagram. Define $\tr(K)=d^{a-n}$. Extend
$\tr$ to all of $\gtl$ by
linearity. 
\end{deff}
For example: 
% of the trace of the Kauffman 4-diagram given
%above:

\[ tr( \vpic {kaufman4diagram.eps} {1in} ) = d^{-4} \vpic
{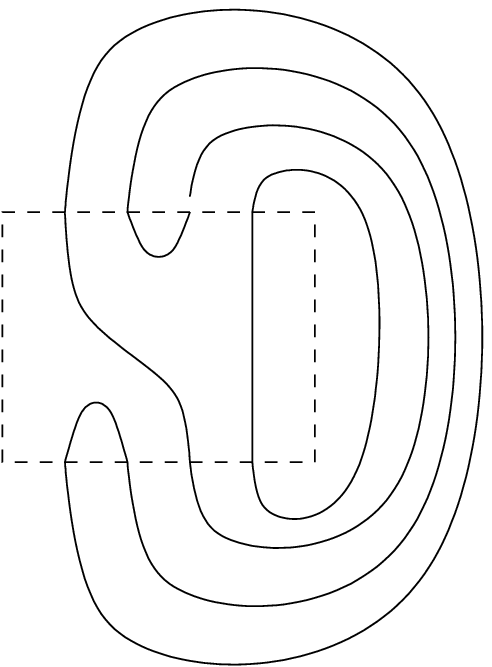} {1.5in} = d^{-2}\]

%We note that the trace map is consistent with the natural
%inclusion of $\gtl \ss gTL_{n+1}(d)$ described earlier.

Since $\tl$ and $\gtl$ are isomorphic, $\tr$ induces a trace 
on $\tl$; for simplicity we shall denote this map 
by $\tr$ as well. 
 
\begin{claim}\label{cl:properties}
$\tr$ satisfies the following three properties: 
\begin{enumerate} 
\item $\tr (1)=1$, 
\item $\tr (XY)=\tr (YX)$ for any $X,Y\in \tl$, 
\item If $X \in TL_{n-1}(d)$ then $\tr (X E_{n-1})=
\frac{1}{d}\tr(X)$. 
\end{enumerate} 
\end{claim}

\begin{proof}
It is straightforward to verify this 
by examining the appropriate pictures in $\gtl$. 
\end{proof}

Of particular importance is the third property, 
which is referred to as the {\it Markov} property. 
These three properties uniquely determine a linear map on $\tl$: 
\begin{lem}\cite{jones:index}\label{lem:markovunique} 
There is a unique linear function $\tr$ on $\tl$ (and on any 
representation 
of it) that satisfies properties 
$1-3$. 
\end{lem} 

\begin{proof}
  By a reduced word $w \in \tl $ we shall mean a word in the set
$\{1, E_1, \dots E_{n-1}\}$
that is not equal to $cw'$ for any $c$ a constant and $w'$ a word of
smaller length.  
Using the relations 
of $\tl$ and applying simple combinatorial arguments, we 
show that a reduced word $w \in \tl$ contains at most one $E_{n-1}$
term. We induct on $n$. 
Clearly the only reduced words in $Tl_2 (d)$ are
$1$ and $E_1$. 
Assume the statement is true for reduced words in $Tl_{n-1}(d)$.
Suppose there exists a
reduced word $w \in \tl$ containing more than one $E_{n-1}$ term. 
Write $w= w_1 E_{n-1} w_2 E_{n-1} w_3$ with $w_2$ a word without
$E_{n-1}$. 
Since $w_2$ must be reduced and is in $Tl_{n-1}(d)$, the induction
hypothesis implies
$w_2$ contains at most one $E_{n-2}$ term. If $w_2$ does not contain
a $E_{n-2}$ term,
$w_2 \in Tl_{n-2}(d)$ and it commutes with $E_{n-1}$ so we have
$w=w_1 w_2 E_{n-1}E_{n-1}w_3$
 which shows that $w$ was not reduced. Otherwise we can write $w_2=
vE_{n-2}v'$ with $v, v'$
both words in $TL_{n-2}(d)$. It follows therefore that $v$ and $v'$
commute with $E_{n-1}$ and
 thus $w= w_1v E_{n-1}E_{n-2}E_{n-1} v' w_3$ which again shows that
$w$ was not reduced.
 We conclude that any reduced word in $\tl$ contains at most one
$E_{n-1}$ term.
 
Given $w \in \tl \backslash Tl_{n-1}(d)$ a reduced word we write
$w=w_1 E_{n-1}w_2$ with $w_1$, $w_2 \in Tl_{n-1}(d)$. Then
$tr(w)=tr(w_2w_1 E_{n-1})=dtr(w_2w_1)$, the first equality by
property 2, the second by property 3. 
Thus for any word $w \in \tl$ we can reduce the trace computation to
the trace of a word $w_2w_1 \in Tl_{n-1}(d)$. Iterating this
process, (and using the fact that $tr(1)=1$), we see that the trace
of a word in $\tl$ is uniquely determined by the relations 1.-3.
Since the trace is linear, the result follows. \end{proof}

We have the following 
convenient description of the Jones polynomial in terms of the Markov
trace. 
\begin{lem}\label{lem:jonestrace}
Given a braid $B$, then 
\[ V_{B^{tr}}(A^{-4})= (-A)^{3w(B^{tr})}d^{n-1} 
tr(\rho_A(B)). \]
\end{lem}

\begin{proof} By Definition
\ref{def:jones}, 
we need to show that $<B^{tr}>=tr(\rho_A(B))d^{n-1}$.
We observe that there exists a one to one 
correspondence between states that appear in the 
bracket sum $<B^{tr}>$, and Kauffman $n$-diagrams that appear in 
$\rho_A(B)$. 
The weight of an element in the bracket state sum corresponding 
to the state $\sigma$ is 
 $A^{\sigma^+-\sigma^-} d^{|\sigma|-1}$. We observe that the
corresponding 
 Kauffman $n$-diagram appears in $\rho_A(B)$ with the
 weight $A^{\sigma^+-\sigma^-}$. Hence, by linearity of the trace, 
it remains to show that for each $\sigma$, 
the trace of the Kauffman diagram corresponding 
to $\sigma$, times $d^{n-1}$, 
equals to the remaining factor in the contribution of $\sigma$ 
to the bracket state sum, $d^{|\sigma|-1}$. 
This is true since by the
 definition of the trace of a Kauffman diagram, it is exactly 
$d^{|\sigma|-n}$. 
 \end{proof}

This lemma also holds if $B$ is replaced by a 
tangle.

%If $B \in B_{2n}$ then
%\[V_{B^{pl}}(A^{-4})= (-A)^{3w(B^{tr})}\frac{1}
%{-(A^2 + A^{-2})} tr(\phi(B)E_1E_3 \dots E_{2n-1}). \]
%\end{lem} 

\subsection{The Path Model Representation of 
$TL_n(d)$}\label{sec:pathmodel}
We describe the path model representation of $TL_n(d)$ due to 
\cite{jones:hecke,jones:poly}.  
\ignore{ As we will see later, 
this representation
induces a unitary representation of the braid group $B_n$, 
 via the representation of $B_n$ inside $TL_n(d)$
 (Definition \ref{def:embed}).} 
The representation will act on a vector space determined by paths on a 
graph.  Specifically, given an integer $k$ ($k$ will be chosen in 
relation to $d$ later), let $G_k$ be the straight line graph with $k-2$ 
segments and $k-1$ vertices:
\[ \hpic {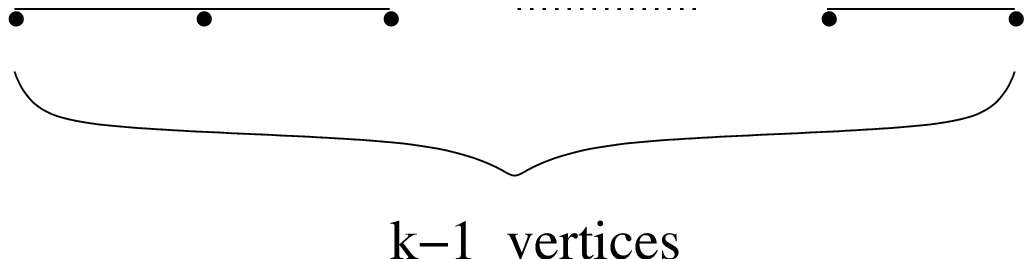} {.7in} \] 

Define $Q_{n,k}$ to be the set of all paths of length $n$ on the graph 
$G_k$ beginning at the leftmost vertex.  Given $q \in Q_{n,k}$, we 
shall denote by $q(0), q(1), \dots q(n)$ the sequence of vertices of $G_k$ 
describing $q$; thus q(0) is the leftmost vertex and $q(i)$ and 
$q(i+1)$ are adjacent vertices of $G_k$ for all $i$.  We shall think of the 
elements of $Q_{n,k}$ as an orthonormal basis of a vector space ${\cal 
V}_{n,k}$; hence an element $q \in Q_{n,k}$ shall represent both a path 
on $G_k$ and a basis element of ${\cal V}_{n,k}$. 
We now construct the path model representation 
$\tau (\tl): {\cal V}_{n,k} \rightarrow  {\cal V}_{n,k}$. 

Given a Kauffman n-diagram $T$, to 
describe $\tau (T)$ it will suffice to give the matrix 
entry $\tau (T)_{q',q}$ 
for each pair $q',q \in Q_{n,k}$.
To do this, we note that the strands of 
a Kauffman diagram separate the rectangle into 
regions; we would like to label the regions by 
vertices of $G_k$, such that the labeling of the 
 bottom part of $T$ will correspond to $q$ and the  top part to $q'$, 
and then compute the matrix element 
$\tau (T)_{q',q}$ from the labels.  
This is done as follows.  

The $n$ marked points of a Kauffman n-diagram 
divide the top and bottom boundary into $n+1$ segments 
which we shall refer to as {\it gaps}. 
We shall say a set of gaps that bound the 
same region in the diagram are {\it connected}.    For example, in the 
following Kauffman 3-diagram:
 \[ \vpic {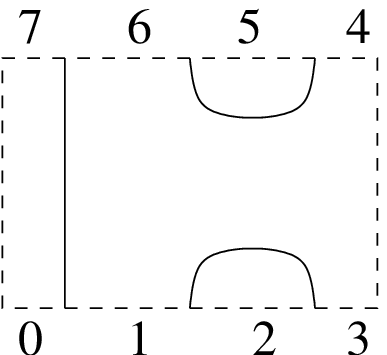} {1in} \]
 the set of gaps $\{ 0,7\}$ are connected as are the set of 
 gaps $\{ 1,3,4,6 \}$.
  We shall 
say that the pair $(q',q)$ is {\it compatible} with $T$ if once we 
label 
the gaps on the bottom from left to right by $q(0), q(1), \dots, q(n)$ 
and we label the gaps on the top from left to right by $q'(0), q'(1), 
\dots , q'(n)$, then any set of 
connected gaps are all labeled by the same 
vertex of $G_k$. Thus in this case 
each region of $T$ can be thought
of as being labeled by a single vertex of $G_k$.

The matrix entry $\tau (T)_{q',q}$ will only be nonzero  
in case the pair of paths $(q',q)$ is {\it compatible} with $T$.  
\ignore{We want 
to assign values to 
matrix entry $\tau(T)_{q',q}$ for compatible $(q',q)$, 
in such a way  that the relations 
in Definition \ref{d:1} are satisfied.}
In this case, the regions are indeed labeled by vertices in $G_k$; we 
can 
now do the following.  
  To each local maximum and minimum of the Kauffman diagram $T$, we 
associate a complex number that depends on the labeling of the 
regions that surround them as follows: \\

 \begin{center}
 $ \hpic {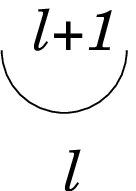} {.4in} $ $\mapsto a_\ell$ , 
$ \hpic {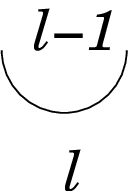} {.4in} $, $\mapsto b_\ell$
 $ \hpic {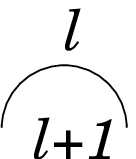} {.4in} $ $\mapsto c_\ell$, 
 $ \hpic {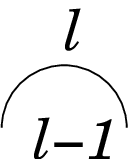} {.4in} $ $\mapsto d_\ell$
\end{center}
The matrix element $\tau(T)_{q',q}$ at a compatible 
pair $(q',q)$, is defined to be
 the product of the appropriate complex numbers over all local 
maxima and minima in $T$.

For the map $\tau(T)$ described above to be well-defined, it has to 
give the same result for isotopic Kauffman diagrams.  
An isotopic move can be seen to only 
create or eliminate local maxima and minima in pairs; we 
see that the conditions
\begin{equation} \label{e:isotopy}
b_{\ell+1}c_{\ell}= 1=a_{\ell-1}d_{\ell}
\end{equation}
are necessary and sufficient for the map to be 
isotopically invariant.  A single extra constraint is 
needed to produce a representation of  $\tl $ :

\begin{claim} 
If the coefficients $a_\ell, b_\ell, c_\ell, d_\ell$ satisfy 
(\ref{e:isotopy}) and
\begin{equation}\label{eq:drep} 
d=b_\ell d_\ell+a_\ell c_\ell, 
\end{equation}
then the resulting map $\tau$ defined as above is  
a representation of $TL_n(d)$ .
\end{claim}
\begin{proof} 
To prove the result we need only verify
that  the matrices $\tau(E_i)$
satisfy the relations of Definition \ref{d:1}. 
This amounts to 
verifying that the matrix elements of the operators 
on both sides of each relation
are equal. Pictorially, a matrix element 
in a product of operators is given by stacking the operators together, 
and summing up over all possible labeling inside loops, such 
that the label inside the loop is different by exactly one from the 
label outside the loop.  
This summation corresponds to the summation over the 
intermediate index in matrix multiplication. 

We now check for the different relations in Definition \ref{d:1}.  
For the generator $E_i$, 
there are only four types of non-zero elements, 
namely, four types of compatible pairs corresponding to the 
following types of labeling of the regions near the $i$-th strand:  
 $\hpic {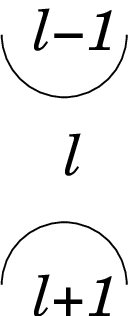} {.7in} $, 
 $\hpic {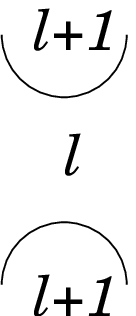} {.7in} $, 
 $\hpic {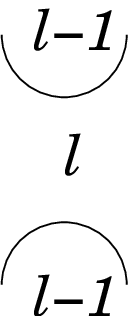} {.7in} $, 
 $\hpic {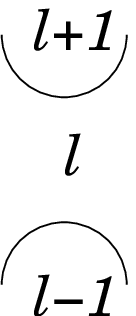} {.7in} $

In the first two relations no loops are created when the operators are 
multiplied, so the verification follows from the 
isotopy invariance of $\tau$ (i.e., Equation (\ref{e:isotopy})). 
The third relation follows from Equation (\ref{eq:drep}),
%$d=b_\ell d_\ell+a_\ell c_\ell$, 
using the fact that one loop was 
created and there are two possible ways to label the region inside.  
%each corresponding 
%to one of the summands. 
\end{proof} 

We would like $\tau(E_i)$ to be Hermitian, 
so that the induced representation on $B_n$ is unitary
by Claim \ref{cl:braidrepunitary}. 
For this we add the constraints 
\begin{equation}\label{eq:conj} 
a_\ell=c_\ell^*, b_\ell=d_\ell^*
\end{equation}

\begin{claim} 
If the coefficients $a_\ell, b_\ell, c_\ell, d_\ell$ satisfy 
Equations (\ref{e:isotopy})-(\ref{eq:conj}) then  
$\tau(E_i)=\tau(E_i)^\dagger$ for all $i$. 
\end{claim} 

\begin{proof} 
We need to prove that
$\tau(E_i)_{q,q'}=(\tau(E_i)_{q',q})^*$.
We need to check only for compatible pairs, namely for $q,q'$ 
which on the $i-1,i,i+1$ entry are equal to 
$\ell,\ell\pm 1,\ell$ for some $\ell$.  
This amounts to checking that 
$a_\ell d_\ell=b_\ell^*c_\ell^*$, $b_\ell d_\ell=b_\ell^* 
d_\ell^*$, 
and $a_\ell c_\ell=a_\ell^* c_\ell^*$. 
This follows from (\ref{eq:conj}). 
\end{proof}

It is now left to solve Equations \ref{e:isotopy} --
\ref{eq:conj} to 
derive 
the definition of $\tau$. 

\begin{claim}\label{cl:solve} 
Define $\lambda_\ell = sin(\pi \ell/k)$ for $\ell\in\{1,..., k-1\}$. 
Then $a_\ell=c_\ell^*=\sqrt{\frac{\lambda_\ell}{\lambda_{\ell-1}}}$, 
$b_\ell=d_\ell^*=\sqrt{\frac{\lambda_{\ell}}{\lambda_{\ell+1}}}$ 
satisfy Equations (\ref{e:isotopy})-(\ref{eq:conj}), 
with $d=2cos (\pi/k)$. 
\end{claim} 

\begin{proof} 
We solve the equations directly. Define 
$x_\ell=\frac{a_\ell}{b_{\ell+1}}$, 
and solve for $x_\ell$. ($x_\ell$ is defined for $\ell\in 
\{1,\dots,k-2\}$). 
We have $\frac{1}{x_{\ell-1}}+x_\ell=d$, and if we set 
$x_0\dots x_\ell=y_\ell$ (and $y_\ell=0$ whenever undefined) 
we have 
$y_{\ell-2}+y_\ell=dy_{\ell-1}$. This is the familiar equation,
defining the eigenvector of the adjacency matrix of the graph $G_k$ 

\begin{equation*}
\left( 
\begin{array}{ccccccc}
0 & 1 & \cdots& & & 0 \\ 1 & 0 & 1 & & & \\
 & 1 & & & & \\
\vdots & & & \ddots & & \\\ & & & &0 & 1 \\ 0& & \cdots& &
1 & 0 \\
\end{array}
\right),
\end{equation*}

with 
eigenvalue $d$.   
It is not difficult to check that 
the $(k-1)$-dimensional vector $\lambda$ is exactly this eigenvector. 
We have $x_\ell=y_\ell/y_{\ell-1}, x_\ell=a_\ell^2$, which gives 
$a_\ell=\sqrt{\lambda_\ell/\lambda_{\ell-1}}$. 
The Equations $b_\ell=1/a_{\ell-1}$ and (\ref{eq:conj}) determine 
$b_\ell, c_\ell, d_\ell$.  
\end{proof}

\dfinalnote{in final version I think we 
should actually explicitly say the resulting matrices, 
 to compare with what we do later when we move to qubits.?}

Using these coefficients, we get the definition of $\tau(E_i)$
as follows.  
$\tau(E_i)_{q,q'}=0$ if $(q,q')$ is not compatible with $E_i$. 
For a compatible pair $(q,q')$, 
$\tau(E_i)_{q,q'}$ is the product of two coefficients, one corresponding 
to the maximum and the other to the minimum in $E_i$.  
For example, if $q=q'$ both move from the site
 $\ell$ to $\ell+1$ in the $i$th step, and then return to 
the site $\ell$ in the $i+1$th step, we have  
$\tau(E_i)_{q,q}=a_\ell c_\ell$. 
Given $\tau(E_i)$, we can extend our definition 
of $\tau$ to all elements in $\tl$. 
    
\subsection{Unitary Path Model Representation of 
$B_n$}\label{sec:unitarypathmodel}
The previous section provided a representation $\tau$ of $\tl$ 
defined for $d=2cos(\pi/k)$.  
We have that
$d=-A^2-A^{-2}$ for $A=ie^{-\pi/2k}$, and also 
that  $\tau(E_i)$ are Hermitian.   
Thus, the conditions of Subsection \ref{sec:unitary} are satisfied. 
We define: 

\begin{deff}\label{deff:varphi} 
The unitary path model representation of $B_n$ is defined to be 
$\varphi(B)=\tau(\rho_A(B))$. 
\end{deff}

The map $\varphi$ can be extended to operate on tangles 
by letting $\rho_A$ be applied only to the crossings in the tangle.

\section{The Quantum Algorithm}\label{sec:alg} 
We are now ready to prove Theorems 
\ref{thm:maintrace} and \ref{thm:mainplat}. 
We first translate the path model representation to work on qubits, 
and show that it can be implemented efficiently. 
We use this to design the algorithms, and then prove their 
correctness. 

\subsection{Moving to Qubits} 
The adaptation of the path model representation to qubits is fairly 
straightforward. We simply switch from presenting paths by the list 
of their locations, to a binary representation which 
indicates the {\it direction} of each step. 
Thus, we shall interpret a string of $n$ bits to be a sequence of 
instructions, where a $0$ shall mean take one step to the left
 and a $1$ shall mean take one step to the right. We shall restrict 
our attention to those $n$ bit strings that describe a path that 
starts at the leftmost vertex of $G_k$ and remains inside $G_k$ at 
each step. From here on when we say ``path'' 
we actually mean the bit string that represents the 
path. 

 \begin{deff}
We define $P_{n,k,\ell}$ to be the set of all paths $p$ on $G_k$ of
$n$ steps 
which start at the left most site and end at the $\ell$'s site. 
We define the subspace $\calH_{n,k,\ell}$  
to be the span of $\ket{i}$ over all $i\in P_{n,k,\ell}$. 
In a similar way, we define $P_{n,k}$ to be all paths with no
restriction 
on the final point, i.e., $P_{n,k} = \cup_{l=1}^{k} P_{n,k,l}$, and we 
define $\calH_{n,k}$ to be the span of the corresponding
computational basis states. 
\end{deff}

We define a representation $\Phi$ as a homomorphism from $TL_n(d)$ 
to matrices operating on ${\cal H}_{n,k}$. 
To define $\Phi$ it suffices to specify the images of the
$E_i$'s, $\Phi(E_i)=\Phi_i$.  
The operators $\Phi_i$ are defined so that 
they correspond to the operators $\tau(E_i)$ (see Subsection 
\ref{sec:pathmodel}) via the natural isomorphism between 
${\cal V}_{n,k}$ and ${\cal H}_{n,k}$. 
Thus, the transition from $\tau$ to 
$\Phi$ is merely a change of language. 
This is done as follows. 

To uniquely define $\Phi_i$ on ${\cal H}_{n,k}$, it suffices to define 
what
it does to each basis element, namely, to $\ket{p}$ for $p\in P_{n,k}$. 
We need the following notation: 
\begin{deff}
Let $p|_i$
denote the restriction of a path $p$ to its first $i-1$
coordinates. 
Given a path $p$ on $G_{k}$, 
we denote by $\ell(p)\in \{1,...,k-1\}$ 
the location in $G_{k}$ that the path $p$ reached in its final
site. 
Denote $z_i=\ell(p|_{i})$. 
\end{deff}

We can now define the operation of $\Phi_i$.
$\Phi_i$ is defined as an operation on the first $i+1$ coordinates 
in a path $p$: 

\begin{eqnarray}\label{deff:phi}
\Phi_i\ket{p|_i00}&=& 0\\\nonumber
\Phi_i\ket{p|_i01}&=& 
\frac{\lambda_{z_i-1}}{\lambda_{z_i}}\ket{p|_i01}+
\frac{\sqrt{\lambda_{z_i+1}\lambda_{z_i-1}}}
{\lambda_{z_i}}\ket{p|_i10}\\\nonumber 
\Phi_i\ket{p|_i10}&=& 
\frac{\lambda_{z_i+1}}{\lambda_{z_i}}\ket{p|_i10}
+\frac{\sqrt{\lambda_{z_i+1}\lambda_{z_i-1}}}{\lambda_{z_i}}
\ket{p|_i01}\\\nonumber
\Phi_i\ket{p|_i11}&=&0\\\nonumber
\end{eqnarray}
 To apply $\Phi_i$ on the $n$-bit string $\ket{p}$
 we tensor the above transformation with identity 
on the last $n-i-1$ qubits. 
For dealing with the edge cases, 
we use the convention $\lambda_j=0$ for any $j\not \in
\{1,..,k-1\}$. 

Once we have defined $\Phi_i$, 
$\Phi$ is then extended to the entire algebra 
by the multiplication property of a representation, and by
linearity.

\begin{claim}\label{cl:rep}
$\Phi$ is a representation of $\tl$. $\Phi_i$ are Hermitian. Moreover, 
$\Phi$ induces a unitary
 representation of the Braid group $B_n$, operating on ${\cal 
H}_{n,k}$. 
\end{claim}

\begin{proof}
The proof follows from the corresponding properties of $\tau$,
namely Subsections \ref{sec:pathmodel}, \ref{sec:unitarypathmodel}, 
and the 
natural isomorphism between paths
on $G_k$ presented by their sequence of locations, ${\cal V}_{n,k}$, 
and paths presented as bit strings, namely ${\calH}_{n,k}$. 
\end{proof}

With the risk of confusion, we denote the unitary representation
of $B_n$ induced by $\Phi$,  
also by $\varphi$ as in Definition \ref{deff:varphi}. 
The only difference is that 
now $\varphi(B)$ operates on ${\cal H}_{n,k}$ rather than
on ${\cal V}_{n,k}$.  

\subsection{Efficient application of one crossing} 
The matrices $\varphi_i$ are defined 
so far only on ${\calH}_{n,k}$ which is a subspace of the Hilbert space 
of 
$n$ qubits; we arbitrarily define their extension to the rest of the 
Hilbert space to be the identity. 

\begin{claim}\label{cl:efficient}
For all $i\in \{1,...,n\}$, $\varphi_i$ can be implemented on the
Hilbert space of $n$ qubits using $\poly(n,k)$ gates.
\end{claim}

\begin{proof}
We note that the application of $\varphi_i$ on $p\in P_{n,k}$ 
modifies only the $i,i+1$ bits of $p$, and the modification depends 
on the location up to the $i$th step, namely on $z_i=\ell(p|_{i-1})$. 
$z_i$ is a number which can be calculated efficiently 
and written on $O(log(k))$ ancilla qubits,
using the following standard technique:
Initialize a counter register of, say, $log(2k)$ qubits to the value 
$1$. Then move along the qubits of 
the path $p$ from left to right, and for each of the $n$ qubits 
update the current state of the counter, $\ell$, by applying 
$$\ket{b}\ket{\ell}\mapsto \ket{b}\ket{\ell+(-1)^b \mod 2k}$$
where $b$ is the state of the currently read qubit. 
Since this is a unitary operation on $log(2k)+1$ 
qubits, it can be applied using polynomially in $k$ many elementary 
quantum gates (this is a standard result in quantum 
computation). 
We end up with the extra register carrying $\ell(p|_{i-1})$. 
 
Now, $\varphi_i$ depends only on the location $\ell(p|_{i-1})$ and on
the 
$i$ and $i+1$ qubits. Hence, once again we have a unitary
transformation 
which operates on logarithmically in $k$ many qubits, and so we 
can implement it in polynomially in $k$ many quantum gates. 
 
After we apply $\varphi_i$, we erase the calculation of
$\ell(p|_{i-1})$ by 
applying the inverse of the first transformation which wrote the
location 
down. 
\end{proof} 

As a corollary, we can deduce that 
\begin{corol}\label{cor:efficient} 
For every braid $B\in B_n$, with $m$ crossings, 
there exists a quantum circuit $Q(B)$ 
that applies $\varphi(B)$ on $n$ qubits, using $\poly(m,n,k)$ 
elementary
gates. 
\end{corol}

\begin{proof}
Order the crossings in the braid in topological order, and apply
the corresponding unitary matrix of each crossing, $\varphi_i$,
one by one, in that order. 
Each crossing takes ${\rm poly}(n,k)$ elementary gates by 
Claim \ref{cl:efficient}, and there are $m$ of them. 
\end{proof}

\subsection{The Algorithms} 
We can now describe the algorithms. The input for both is 
a braid of $n$ strands and $m$ crossings, and an integer $k$. 

\noindent-----------------------------------------------------------------------

\noindent {\bf {\sf Algorithm Approximate-Jones-Plat-Closure}} 
\begin{itemize}
\item Repeat for $j=1$ to $poly(n,m,k)$: 
\begin{enumerate} 
\item Generate the state $|\alpha\ra=|1,0,1,0, \ldots, 1,0\ra$ 
\item Output a random variable $x_j$ whose expectation value is 
${\cal R}e\la \alpha | Q(B)|\alpha\ra$ using the $\hadamard$ test. 
\end{enumerate} 
\item Do the same but for random variables $y_j$ whose expectation 
value is 
${\cal I}m\la \alpha | Q(B)|\alpha\ra$ using the appropriate  
variant of the $\hadamard$ test.

\item Let $r$ be the average over all $x_j+iy_j$ achieved this way. 
Output $(-A)^{3w(B^{pl})}d^{3n/2-1}\lambda_1 r/ N$. 
\end{itemize}

\noindent-----------------------------------------------------------------------

\noindent {\bf {\sf Algorithm Approximate-Jones-Trace-Closure}} 
\begin{itemize} 
\item 
Repeat for $j=1$ to $poly(n,m,k)$: 
\begin{enumerate}
\item Classically, pick a random path $p\in P_{n,k}$ with probability 
$Pr(p)\propto\lambda_\ell$, where $\ell$ is the index of the site which 
$p$ ends at. 
\item Output a random variable $x_j$ whose expectation value is 
${\cal R}e\la p | Q(B)|p\ra$ using the $\hadamard$ test. 
\end{enumerate}

\item Do the same but for random variables $y_j$ whose expectation 
value is 
${\cal I}m\la p | Q(B)|p\ra$ using the appropriate  
variant of the $\hadamard$ test.

\item Let $r$ be the average over all $x_j+iy_j$. 
Output \newline $(-A)^{3w(B^{tr})}d^{n-1}r$. 
\end{itemize}

\noindent-----------------------------------------------------------------------

\begin{claim} 
The above two quantum algorithms can be performed in time 
polynomial in $n,m,k$. 
\end{claim}

\begin{proof}
Algorithm {\sf Approximate-Jones-Plat-Closure}
 is clearly efficient, because 
 the $\hadamard$ test can be applied efficiently using 
 Corollary \ref{cor:efficient}. 
To perform the first step of the second algorithm 
 efficiently, we pick
 a random $\ell\in \{1,\ldots,k\}$, with probability 
proportional to $\lambda_\ell$, and then use \znote{changed wording for 
space}
%the following Claim \ref{cl:random}. 
Claim \ref{cl:random} below.
We note that our calculations involve irrational numbers, 
$\lambda_\ell$, but these can be approximated to within an 
exponentially good precision efficiently. \end{proof} 

\begin{claim}\label{cl:random}
Given $n,k,\ell$, 
there exists a classical probabilistic algorithm which runs
in time polynomial in $n$ and $k$, and
outputs a random path in $P_{n,k,\ell}$ according to a distribution 
which is exponentially close to uniform. 
\end{claim}

\begin{proof}
(We thank O. Regev for a discussion 
that lead to this variant.)
\znote{removed citation to private comm.}
%\cite{regev}
We briefly sketch the proof.
%{\bf Sketch:}     
We first note that the following $n \times k$ array $S$, defined by 
 $S_{i,j}=|P_{i,k,j}|$ (the number of path on $G_k$ of $i$ steps that
start at $1$ and end 
at $j$) can be easily calculated efficiently using the 
recursion $ |P_{n,k,\ell}|=|P_{n-1,k,\ell-1}|+|P_{n-1,k,\ell+1}|.$
To pick a random path ending at $\ell$, we pick the values 
of the sites in $p$ one by one in reverse order. 
If $p(j)=\ell$, we randomly decide if $p(j-1)=\ell-1$ or $\ell+1$ 
according to the ratio $S_{j-1,\ell-1}:S_{j-1,\ell+1}$. Given 
$p(j-1)$ we can continue 
to pick $p(j-2)$ in a similar manner, and so on. 
\end{proof}

\begin{thm}\label{thm:tracecorrect} 
Algorithm {\sf Approximate-Jones-Trace-Closure} approximates the 
Jones polynomial of $B^{tr}$ at $A^{-4}=e^{2\pi i/k}$, 
to within the precision specified in Theorem \ref{thm:maintrace}. 
\end{thm} 

\begin{proof} 
We will need the following definition. 
 
\begin{deff}\label{deff:trn} 
Define $Tr_{n}(W)$ for every $W$ in the image of
$\Phi(TL_n(d))$
to be: 
$$Tr_{n}(W)= \frac{1}{N}\sum_{\ell=1}^{k-1} \lambda_\ell
Tr(W|_\ell)$$ 
where $W|_{\ell}$ denotes the restriction of $W$ 
to the subspace ${\cal H}_{n,k,\ell}$, and 
Tr denotes the standard trace on matrices. 
The renormalization is $N=\sum_\ell \lambda_\ell dim({\cal H}_{n,k,l})$
where 
the sum is taken over all $\ell$'s such that $P_{n,k,\ell}$ is non
empty. 
\end{deff}

This definition makes sense because 
 matrices in the image of $\varphi$
are block diagonal, with the blocks indexed by $\ell$, 
the last site of the paths: 
 
\begin{claim}\label{cl:reducible}
For any $T\in TL_n(d)$,
 $\Phi(T){\cal H}_{n,k,\ell}\subseteq {\cal H}_{n,k,\ell}$. 
\end{claim} 

\begin{proof}
$\Phi_i$ cannot change the final point of a
path since it only moves $01$ to $10$ and vice versa. 
\end{proof} 

Hence, the above trace function simply gives different weights 
to these blocks (and gives zero weights on strings that aren't 
paths). We claim that $Tr_n$ is a Markov trace. 

\begin{claim}\label{cl:markov}
The function $Tr_n(\cdot)$ satisfies the three properties in 
Claim \ref{cl:properties}. 
\end{claim}

\dfinalnote{I'd recommend getting rid of this. . . if you don't, there 
are a bunch of things to do. . . fix the displayed equations that leak 
into the margin, and fix the equations. .. a lot of expressions should 
end with path labelled p1 whereas you use p0 twice}
\begin{proof}
That $Tr_n(\Phi(1))=1$ follows from the renormalization. 
The second property follows from Claim 
\ref{cl:reducible} plus the fact that the standard trace 
on matrices satisfies this property, so $Tr_n(\cdot)$ satisfies it on 
each block separately. To show the Markov property, we
 have to show that if $X \in TL_{n-1}(d))$ then 
$Tr_n (\Phi(X) \Phi(E_{n-1}))= \frac{1}{d}Tr_n(\Phi(X))$.
We note that for any $X \in TL_{n-1}(d)$, $\Phi(X)$ can be
written 
as a linear combination of terms of the form $\ket{p}\langle
p'|\otimes I$, 
with $p,p'\in P_{n-1,k}$ and the identity operates on the last
qubit. 
By linearity, it suffices to prove the Markov property on such
matrices. 
Writing $\ket{p}\bra{p'}\otimes
I=\ket{p0}\bra{p'0}+\ket{p1}\bra{p'1}$
we require: 
 $Tr_n (\ket{p0}\bra{p'0}\Phi_{n-1}+\ket{p1}\bra{p'1}\Phi_{n-1})= 
\frac{1}{d}Tr_n(\ket{p0}\bra{p'0}+\ket{p1}\bra{p'1}).$
This can be easily verified using the definition of $\Phi$ 
 by checking the two cases $p=p'$ and $p\not = p'$. 

We start with the case $p\not = p'$. 
In this case the right hand side is $0$. 
As for the left hand side, 
$\bra{p'0}\Phi_{n-1}$ has a zero component on $\bra{p0}$. To see
this, 
we check the two cases: if $p'$ ends with $0$ 
then $\bra{p'0}\Phi_{n-1}=0$, otherwise $p'$ ends with $1$. 
 $\bra{p'0}\Phi_{n-1}$ is then a sum of 
two terms, one equals to $\bra{p'0}$ and is therefore different
than 
 $\bra{p0}$, and the other ends with $01$ and is thus also
different 
from $\bra{p0}$. The same argument works to show that 
$\bra{p'1}\Phi_{n-1}$ has a zero component on $\bra{p1}$.
Hence, the left hand side is also $0$. 
 
It is left to check the equality in the case $p=p'$.
We require 
 $$Tr_n (\ket{p0}\bra{p0}\Phi_{n-1} +\ket{p1}\bra{p1}\Phi_{n-1})= 
\frac{1}{d}Tr_n(\ket{p0}\bra{p0}+\ket{p0}\bra{p0}).$$
Suppose $\ell(p)=\ell.$ Then the right hand side is equal to 
$\frac{1}{d}(\lambda_{\ell-1}+\lambda_{\ell+1})=\lambda_\ell$, 
using the properties of the eigenvector $\lambda$, as in 
Claim \ref{cl:solve}. 
To see that the left hand side is the same, 
we again divide to cases. 
Suppose first that $p$ ends with $0$. In this case 
$\bra{p0}\Phi_{n-1}=0$. As for the other term, 
 $\bra{p1}\Phi_{n-1}=\frac{\lambda_{\ell}}{\lambda_{\ell+1}}
\bra{p1}$, using the definition of $\Phi_{n-1}$ and the 
fact that $p$ without its last step ends in $\ell+1$. 
The weight in the trace of the left hand side is
$\lambda_{\ell+1}$,
and so the left hand side is equal to $\lambda_\ell$ too. 
The argument is similar in the case that $p$ ends with $1$.
\end{proof} 

By the uniqueness of the Markov trace, Lemma \ref{lem:markovunique}, 
we have that 
$Tr_n(\varphi(B))=tr(\rho_A(B))$. 
Hence, using Lemma \ref{lem:jonestrace},
we have that for any braid $B\in B_n$ 
\begin{lem}\label{le:jonesmatrixtrace} 
$$ V_{B^{tr}}(A^{-4})= (-A)^{3w(B^{tr})}d^{n-1}Tr_n(\varphi(B)).$$
\end{lem} 

Due to Lemma \ref{le:jonesmatrixtrace}, 
the correctness of the algorithm follows trivially 
from the following claim. 

\begin{claim} 
With all but exponentially small probability, 
the output $r$ satisfies 
$|r-Tr_n(\varphi(B))|\le \epsilon$
for $\epsilon$ which is inverse polynomial in $n,k,m$. 
\end{claim} 

\begin{proof}
The $\hadamard$ test indeed implies that 
the expectation of $x_j$ for a fixed $p$ is exactly 
${\cal R}e\la p|\varphi(B)|p\ra$. 
The expectation of the variable $x_j$ taken over a random $p$ 
is thus 
$$\frac{\sum_{\ell, p\in P_{n,k,\ell}} \lambda_\ell
 {\cal R}e(\la p|\varphi(B)|p\ra)}
{\sum_{\ell,p\in P_{n,k,\ell}} \lambda_\ell}=
\frac{\sum_{\ell} \lambda_\ell {\cal R}e(Tr(\varphi(B)|_\ell))}{
\sum_\ell \lambda_\ell dim(H_{n,k,l})}=$$ $$={\cal R}e 
(Tr_n(\varphi(B))).$$ 
The same argument works for the imaginary part. 
Since $r$ is the sum of two averages of polynomially many $i.i.d$ 
random variables, each taking values between $1$ and $-1$, 
the result follows by the Chernoff-Hoeffding bound. 
\end{proof} 

This completes the proof of Theorem \ref{thm:tracecorrect}. 
\end{proof}

\begin{thm}\label{thm:platcorrect} 
Algorithm {\sf Approximate-Jones-Plat-Closure}
 approximates the 
Jones polynomial of $B^{pl}$ at $A^{-4}=e^{2\pi i/k}$, 
to within the desired precision as in Theorem \ref{thm:mainplat}. 
\end{thm}

\begin{proof} 
By the correctness of the $\hadamard$ test, and by the 
Chernoff-Hoeffding bound, 
the variable $r$ which the algorithm computes 
 is, with exponentially 
good confidence, within $\delta=1/poly(n,m,k)$ from 
$\la \alpha|\varphi(B)|\alpha\ra$, which is equal to 
$Tr(\varphi(B)|\alpha\ra\la\alpha|)$. 
We need to connect this expression to the Jones polynomial 
of the plat closure of $B$. The main observation here is that  
the plat closure of a braid $B$ is
isotopic to the trace closure of a tangle $C$ achieved by applying 
 the braid on $n/2$ capcups, as in the following 
picture:

\[ \vpic {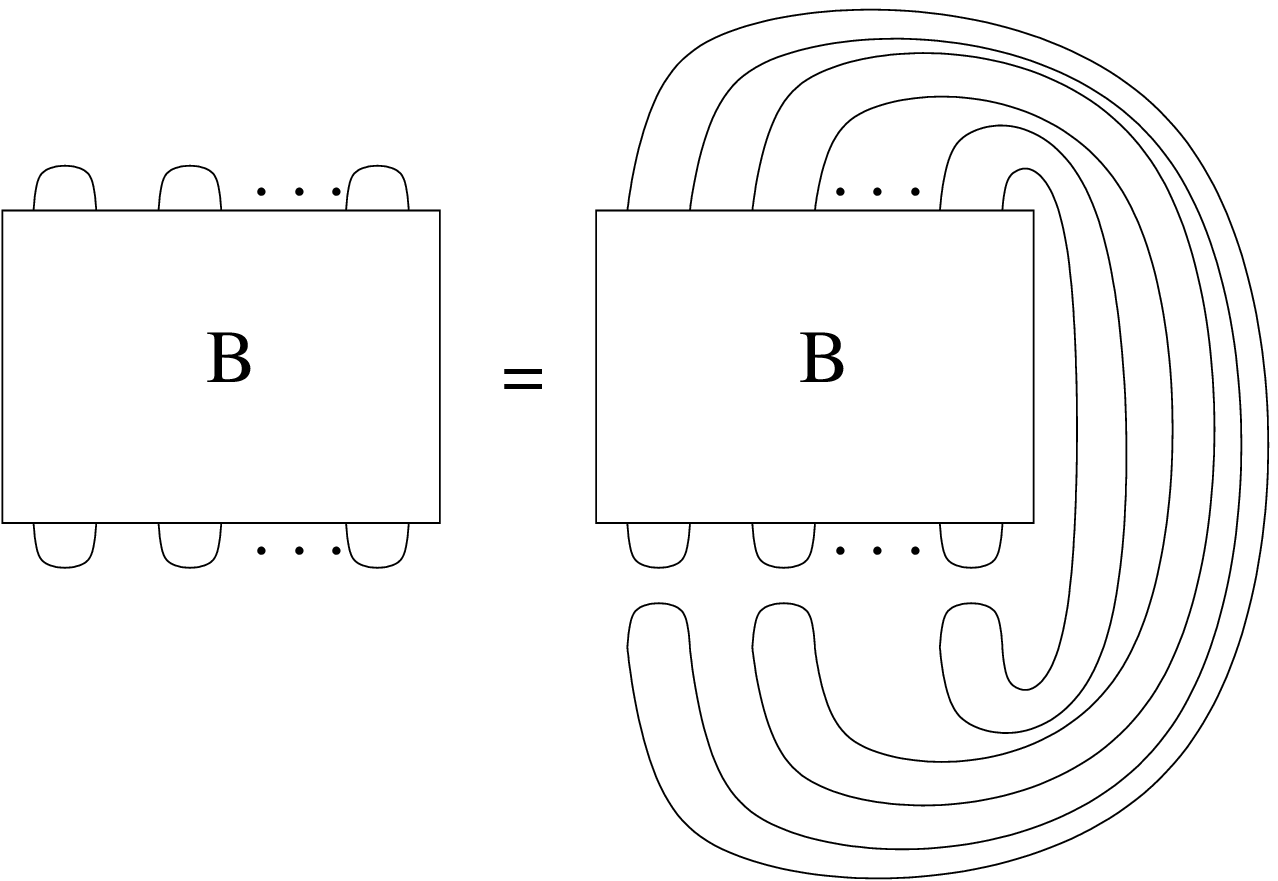} {2in} \]

It therefore suffices to relate  
the Jones polynomial of $C^{tr}$ to 
$Tr(\varphi(B)|\alpha\ra\la\alpha|)$. 
Since the question is now stated in terms of 
trace closures and traces, there is hope to 
be able to apply  Lemma \ref{lem:jonestrace} and Claim \ref{cl:markov} 
as in the proof of 
Theorem \ref{thm:tracecorrect}. 
But we first need to make the connection between 
 the projection on $|\alpha\ra$ and capcups.  
\begin{claim}\label{cl:platcupcap} 
$|\alpha\ra\la\alpha|=\Phi_1\Phi_3\ldots\Phi_{n-1}/d^{n/2}.$
\end{claim} 

\begin{proof} 
It is easy to verify that $\Phi_1\Phi_3\ldots\Phi_{n-1}$ 
applied to any path except for $|\alpha\ra$ gives $0$, 
and when applied to $|\alpha\ra$ it gives the desired factor. 
To do this we use the fact that $\Phi_i$ commute 
if their indices are more than one apart, and so we can first apply 
$\Phi_1$, then $\Phi_3$ and so on. 
 Since the path starts at the left most site, $\Phi_1$ on $p$ 
 simply applies the following rescaled projection: 
$d|10\ra\la10|$ on the first two coordinates. 
This projection forces the first two coordinates to be $10$, 
and so $\Phi_1 |p\ra$ returns to the starting point after two 
steps. Therefore a similar argument applies when we apply
 $\Phi_3$ on the next two coordinates, and so on. 
By induction, we get the desired result. 
\end{proof} 

We thus have, using Definition \ref{deff:trn} and Claim 
\ref{cl:platcupcap}: 
\begin{equation*} 
\la\alpha|\varphi(B)|\alpha\ra=Tr(\varphi(B)|\alpha\ra\la\alpha|)=
\frac{N}{\lambda_1}Tr_n(\varphi(B)|\alpha\ra\la\alpha|)=
\end{equation*}
\begin{equation*}\label{eq:plat} 
=\frac{N}{\lambda_1}
Tr_n(\varphi(B)\Phi_1\Phi_3\ldots\Phi_{n-1}/d^{n/2})= 
\frac{N}{\lambda_1 d^{n/2}}
Tr_n(\varphi(C))
\end{equation*}

By the uniqueness of the Markov trace, Lemma \ref{lem:markovunique}, 
and by Claim \ref{cl:markov}, 
we have that $Tr_n(\varphi(C))=tr(\rho_A(C))$.  
Using Lemma \ref{lem:jonestrace}, 
we have: 
\[ V_{B^{pl}}(A^{-4})=V_{C^{tr}}(A^{-4})= (-A)^{3w(C^{tr})}d^{n-1} 
Tr_n(\varphi(c)).\]
which we substitute in the previous equation. 
This completes the proof of Theorem \ref{thm:platcorrect}.
\end{proof} 

\ignore{
We get that the 
quantity 
$\la\alpha|\varphi(B)|\alpha\ra$ which the algorithm approximates
relates to the Jones polynomial of the plat closure of the braid with 
the 
correct factor, and the approximation is indeed as required.}

%\begin{description}
%Run a simple random walk of length $n$, which 
%except that the next step of the walk at sites $1$ and $k$ is 
%deterministically determined to be $2$ or $k-1$, respectively. 
%Accept the output path it only if 
%it ends at $\ell$. If it doesn't, repeat until you get a walk 
%that ends at $\ell$, or until you have tried for 
%poly(n) many times and failed. If this is the case, then 
%output some arbitrary path that ends at $\ell$. 
%\end{description}
%If the random walk has mixed, then the final state is 
%$\ell$withprobability 
%$1/poly(k)$. Hence, as long as $n$ is $Omega(k^2)$, we are OK. 
%If we only allow to try polynomially many attempts, The result is 
%a random path in by running a random walk of length 
%$n$ on the path of length $k$, and throwing away all 
%the path unless it ends at $\ell$. Because there is fast mixing, 
% and conditioning on the fact that the final point is $\ell$.)

%\bibliographystyle{ieee}
%\bibliography{adbib}
%\begin{thebibliography}{99}

\section{Acknowledgements} 
\noindent \znote{changed to A. Kitaev and U. Vazirani}
We thank Alesha Kitaev for clarifications regarding the 
difference between the plat and trace closure, and the nature of 
the path model representation. 
We are grateful to Umesh Vazirani for helpful remarks regarding the 
presentation.

\end{document}